\documentclass[12]{article}
\usepackage{amsmath,amssymb,amsthm,blkarray}
\usepackage{mathtools}
\usepackage{mathrsfs}
\usepackage{bm}
\usepackage{slashed} 
\usepackage{multirow}
\usepackage{tikz}
\usepackage[utf8]{inputenc}
\usepackage{fullpage}
\usepackage{arydshln}
\usepackage{setspace}
\usepackage[margin=1in]{geometry}
\usepackage{dsfont}
\usepackage{physics}
\usepackage{graphicx }
\graphicspath{ {graphs/} }
\usepackage{subcaption}
\usepackage{float}
\usepackage{xcolor}
\usepackage{cite}
\usepackage{subcaption}
\usepackage{mwe}
\usepackage[toc,page]{appendix}
\usepackage{hyperref}
\hypersetup{colorlinks=true,linkcolor=blue,anchorcolor=blue,citecolor=green,filecolor=blue
urlcolor=blue,bookmarksnumbered=true,pdfview=FitB}

\renewcommand*{\thefootnote}{\fnsymbol{footnote}}
\usepackage{qcircuit}

\usepackage{epstopdf}

\begin{document}
\begin{center}
{\Large \bf \strut Quantum Simulation of Nuclear Inelastic Scattering
\strut}\\
\vspace{5mm}
\today

\vspace{5mm}
{\large 
Weijie Du$^{a,b,c,}$\footnote{duweigy@gmail.com}, James P. Vary{$^b$}, Xingbo Zhao$^{a,c}$, and Wei Zuo$^{a,c}$
}  

\noindent
{\small $^a$\it Institute of Modern Physics, Chinese Academy of Sciences, Lanzhou 730000, China} \\
{\small $^b$\it Department of Physics and Astronomy, Iowa State University, Ames, Iowa 50010, USA  \\
{\small $^c$\it School of Nuclear Science and Technology, University of Chinese Academy of Sciences, Beijing 100049, China} \\
}

\end{center}

\setcounter{footnote}{0} 
\renewcommand*{\thefootnote}{\arabic{footnote}}

\section*{Abstract}
We present a time-dependent quantum algorithm for nuclear inelastic scattering in the time-dependent basis function on qubits approach. This algorithm aims to quantum simulate a subset of the nuclear inelastic scattering problems that are of physical interest, in which the internal degrees of freedom of the reaction system are excited by time-dependent external interactions. 
We expect that our algorithm will enable an exponential speedup in simulating the dynamics of the subset of the inelastic scattering problems, which would also be advantageous for the applications to more complicated scattering problems.    
For a demonstration problem, we solve for the Coulomb excitation of the deuteron, where the quantum simulations are performed with IBM Qiskit.

\section{Introduction}
Following Richard Feynman’s original idea of simulating quantum dynamics using another well-controlled quantum system \cite{Feynman:1981tf}, the last couple of decades have witnessed the exciting innovation and rapid development of quantum computing technology. Quantum computers take advantage of quantum principles to potentially outperform their classical counterparts by increasing the computing power and reducing the required computing resources \cite{Chuang:2000}. Widely celebrated algorithmic innovations such as the Shor’s algorithm \cite{Shor:1997} and the quantum Fourier transformation \cite{Chuang:2000,Coppersmith:1994,Ekert:1996,Camps:2020} fuel the excitement by proving dramatic exponential speedup in computation compared to the classical computers. To exploit the power of the quantum computer, researchers have since developed various algorithms for Hamiltonian dynamics \cite{Lloyd:1996,Childs:2004,Berry:2007,Childs:2011,Poulin:2011,Berry:2012,Wu:2002,Sun:2020}, with applications to problems in, e.g., condensed matter physics \cite{Macridin:2018}, quantum chemistry \cite{Lanyon:2010,Babbush:2015,Colless:2018,McArdle:2020}, nuclear physics \cite{Dumitrescu:2018njn,Roggero:2018hrn}, and quantum field theories \cite{Klco:2018kyo,Klco:2018zqz,Alexandru:2019ozf,Lamm:2019bik,Kreshchuk:2020dla,Kreshchuk:2020kcz,Kreshchuk:2020aiq}. 

Here we present the time-dependent basis function on qubits (tBFQ) approach  for quantum simulating a subset of nuclear inelastic scattering problems, where the internal degrees of freedom of the reaction system are excited by the external interaction. We will also discuss the quantum advantage the tBFQ algorithm would achieve over classical algorithms for this subset of nuclear inelastic scattering problems. With this advantage, the tBFQ algorithm would enable simulations of complicated nuclear inelastic scattering processes that classical computers would find intractable. This treatment is adaptable for atomic and molecular quantum scattering processes \cite{Bates:2012} as well. 

In this work, we will show 1) how to prepare and to qubitize the basis representation; and 2) how to design the time-dependent quantum algorithm for simulating the nuclear inelastic scattering. In particular, we adopt the Schr\"odinger picture and divide the full Hamiltonian of the nuclear system undergoing inelastic scattering into the reference and external interaction parts. We first solve the eigenbasis set of the reference Hamiltonian via nuclear structure calculations on classical computers. Our eigenbases may be obtained from {\it ab initio} methods \cite{Barrett:2013nh, Navratil:2007we,Navratil:2000ww,Navratil:2000gs,Carlson:2014vla}, but eigenbases from phenomenological models would also serve our purposes. We then apply an importance truncation scheme to regulate the basis size. We will calculate the full Hamiltonian and the time-evolution operator using this trimmed basis representation. Next, we qubitize the basis representation. Based on the matrix of the time-evolution operator in the basis representation, we apply the Trotterization technique to decompose the total evolution into a sequence of short-time evolutions \cite{Wu:2002,Sun:2020} (see Eq. \eqref{eq:productFormula} below): 1) this sequence of evolutions is parameterized by the discretized scattering time; and 2) for each time step, the evolution operator of the reaction system is considered as time-independent. We also introduce our idea to design the quantum circuit such that it is directly parameterized by the scattering time (see in Sec's \ref{sec:Sec3p4} and \ref{sec:CircuitConstruction}). Finally, we simulate the time-dependent inelastic scattering process on quantum computer, where the measurements produce the transition probabilities as well as other observables of the reaction system.

We demonstrate the tBFQ algorithm with a simple model problem: the Coulomb excitation of the deuteron in a peripheral scattering with a heavy ion. This specific model problem was applied, in Refs. \cite{Du:2018tce,Du:2017ckx}, to demonstrate the time-dependent basis function (tBF) approach on classical computers, which was introduced to achieve a unified description of nuclear structure and reactions. Though intuitively simple, this model problem is a non-trivial problem due to the strong time-dependent external field that features higher-order transitions to states not directly accessible from the initial state. A generalization of this initial application has successfully reproduced experimental results \cite{Yin:2019kqv}. Here, we similarly benchmark the tBFQ algorithm with this model problem applying the IBM Qiskit quantum simulator \cite{Santos:2017,Cross:2017}. 

The arrangement of this paper is as follows. In Sec. \ref{sec:theory}, we discuss the elements of the scattering theory related to this work. In Sec. \ref{sec:Algorithm}, we present the details of the tBFQ algorithm. We illustrate the model problem and the simulation conditions in Sec's. \ref{sec:ModelProblem} and \ref{sec:simulationConditions}, respectively. In Sec. \ref{sec:resultsAndDiscussions}, we show the results of the model problem via quantum simulations, where we also compare the results of the tBFQ with those of the tBF approach to validate the quantum simulation results. We conclude with Sec. \ref{sec:FinalSec}, where we also provide the outlook. We present in the Appendix a preliminary prescription to decompose a diagonal unitary.

\section{Theory}
\label{sec:theory}
In this work, we focus on the Hamiltonian dynamics of a subset of the nuclear inelastic scattering problems in which the internal degrees of freedom of the reaction system are excited by strong time-dependent external interactions. In this section, we show our formalism of the Hamiltonian dynamics and define our problem set. We also present the details of the construction of the basis representation. 

\subsection{Hamiltonian dynamics}
\label{sec:Hdynamics}
In the Schr\"odinger picture, the equation of motion of the reaction system is described by the time-dependent Schr\"odinger equation\footnote{In this work, we adopt the natural units and take $\hbar = c = 1$. For notation, we use $\hat{U}$, $U$, $\widetilde{U}$ to denote the operator, its matrix representation, and the corresponding gate in the quantum circuit, respectively.}
\begin{align}
i \frac{\partial}{\partial t} |\psi ; t \rangle = \hat{H}(t) |\psi ; t \rangle , \label{eq:SchrodingerEq}
\end{align}
where the full Hamiltonian of the reaction system, $\hat{H}(t) $, can be divided into two self-adjoint terms:
\begin{align}
\hat{H}(t) = \hat{H}_0 + \hat{V}_{\rm int}(t) , \label{eq:RefAndIntHamiltonians}
\end{align} 
with $\hat{H}_0$ denoting the reference Hamiltonian (assumed to be time independent) that determines the available excitations of the reaction system. $\hat{V}_{\rm int}(t)$ is the time-dependent external interaction that drives the dynamic excitation processes.

In this work, we focus on a subset of the nuclear inelastic scattering problems. For such problems, we assume that $ \hat{V}_{\rm int} (t) $ can be divided according to, e.g., different sources, and/or characteristics (e.g., time structure and scale):
\begin{align}
\hat{V}_{\rm int} (t)  = \sum ^m_{i=1} \hat{V}_{i} (t) , \label{eq:summationOfInteraction}
\end{align}
where $i=1,\ 2, \ \cdots , \ m$ is the summation index. We take the total number of such terms to be $m$. We note that, for general applications in nuclear reaction theory, $m$ is determined by a few key characteristics of possible reaction channels. As only a limited number of such characteristics (e.g., symmetries) control the reaction dynamics, we expect $m$ to be a small number (e.g., $m=2 $ would be sufficient for the model problem discussed in Sec. \ref{sec:ModelProblem} as well as its extension to a realistic scattering problem presented in Ref. \cite{Yin:2019kqv}). Furthermore, we assume that each term $\hat{V}_i(t)$ can be factorized into two parts: the time-dependent external part $a_i(t)$ (e.g., due to the external time-varying fields), and the time-independent part $\hat{W}_i$ that characterizes the intrinsic properties of the reaction system
\begin{align}
\hat{V}_i(t)  = a_i(t) \hat{W}_i \label{eq:factorizationOfInteractions}.
\end{align}

The state vector of the system at $t \geq t_0 $ can be solved based on Eq. \eqref{eq:SchrodingerEq} as 
\begin{align}
|\psi ; t \rangle = \hat{U}(t;t_0) |\psi ; t_0 \rangle = \hat{T} \Bigg\{ \exp \Big[ -i \int _{t_0}^t \big( \underbrace{ \hat{H}_0 + \hat{V}_{\rm int}(t') }_{\equiv \hat{H}(t')} \big) dt' \Big] \Bigg\} |\psi ; t_0 \rangle , \label{eq:timeEvolutionOperator}
\end{align}
where $\hat{U}(t;t_0)$ is the time-evolution operator that propagates the system from $t_0$ to $t$. $\hat{T}$ is the time-ordering operator.

\subsection{Basis representation}
\label{sec:BasisRepresentation}
For a full Hamiltonian of $N_d$ degrees of freedom (or dimension of the Hamiltonian in a matrix representation), quantum computers take $poly(N_d, \Delta t)$ number of gates to simulate the time-evolution operator for a given time increment $\Delta t$. For practical calculations, it is important to select a representation that efficiently accounts for the essential degrees of freedom of the full Hamiltonian. While the lattice representation \cite{Klco:2018zqz,Kogut:1979wt} and the particle number representation (see, e.g., Refs. \cite{Ortiz:2002,McArdle:2020}) could also work in reducing the dimension of reaction problems, we suggest that the basis representation is advantageous because of its capacity to incorporate both the bound and scattering channels of nuclear systems \cite{Yin:2019kqv,Johnson:2019sps}. 

The construction of a basis representation suitable for the initial applications that we contemplate has been detailed in our previous works \cite{Du:2018tce,Du:2017ckx,Yin:2019kqv}. In particular, we solve the eigenequation of a reference Hamiltonian $\hat{H}_0$ [Eq. \eqref{eq:RefAndIntHamiltonians}] of the reaction system:
\begin{align}
\hat{H}_0 |\beta _j \rangle = E_j |\beta _j \rangle \label{eq:EigenEquationH0},
\end{align}
where $E_j$ denotes the eigenenergy corresponding to the eigenvector $| \beta _j \rangle$. The subscript ``$j$'' is an index which runs over individual state vectors. The eigenbasis set $\{ | \beta _j \rangle \}$ is then adopted to construct the basis representation for the reaction system that will be subject to the external interaction characterized by $\hat{V}_{\rm int}(t)$. In principle, the basis set has infinite dimension so approximations based on the details of the application enter into consideration. For our test application, we apply an importance truncation scheme to regulate the basis size while retaining the predominant inelastic scattering physics (see, e.g., an example in Ref. \cite{Yin:2019kqv}, where we verified the independence of the obtained results from the basis parameters and from the basis cutoff).

Within the basis representation, we evaluate the matrix element of the reference Hamiltonian $\hat{H}_0$ in Eq. \eqref{eq:RefAndIntHamiltonians} as
\begin{align}
 \langle \beta _j | \hat{H}_0 | \beta _k \rangle = E_j \delta _{jk} . \label{eq:H0MatElement} 
\end{align}
The basis set $\{ |\beta _j \rangle \}$ constructed from $\hat{H}_0$ characterizes only the intrinsic degrees of freedom of the reaction system, and the matrix element of $\hat{V}_{\rm int}(t)$ in the basis representation admits the form:
\begin{align}
\langle \beta _j | \hat{V}_{\rm int} (t) | \beta _k \rangle = \sum ^m_{i=1} a_i(t) \ \langle \beta _j | \hat{W}_i | \beta _k \rangle ,
\end{align}
where we have applied the assumptions of additivity and factorization [Eqs. \eqref{eq:summationOfInteraction} and \eqref{eq:factorizationOfInteractions}].

The state vector of the reaction system in the basis representation is
\begin{align}
|\psi ; t \rangle = \sum _{j} C_{j}(t) |\beta _j \rangle , \label{eq:fullyEntangledWF}
\end{align}
with the amplitude $C_{j} (t) \equiv \langle \beta _j | \psi ; t \rangle $ corresponding to the basis $|\beta _j \rangle $. 

In general, the expectation value for the operator $\hat{O}$ representing an observable of the reaction system can be expressed using the fully entangled state vector of the reaction system as
\begin{align}
\langle \psi ; t | \hat{O} | \psi ; t \rangle  = \sum _{j,k} {C_j}^{\ast}(t) C_k(t) \langle \beta _j | \hat{O} | \beta _k \rangle . 
\end{align}
In this work, we limit our discussion to scalar operators $\hat{O}$, for which $ \langle \beta _j | \hat{O} | \beta _k \rangle = 0 $ if $j \neq k$ in our chosen basis representation. This enables us to avoid evaluating the interference terms (with $j\neq k$), which requires additional research efforts beyond the scope of the current work. An experiment corresponding to this operator makes a projection onto an available state in the basis space of the reaction system leading to a measurement of the expectation value $ \langle \hat{O} (t) \rangle $ according to the formalism of the ensemble average:
\begin{align}
\langle \hat{O}(t) \rangle = \sum _j p_j (t) \langle \hat{O} \rangle _j , \label{eq:observable}
\end{align}
where $ p_j (t) \equiv |C_{j}(t)|^2 $ and $ \langle \hat{O} \rangle _{j} \equiv \langle \beta _j | \hat{O} | \beta _j \rangle $. As in Ref. \cite{Du:2018tce}, we remark that the asymptotic value $ \langle \hat{O}(t_f) \rangle $ (at the end of the scattering when the external field is sufficiently weak) is the quantity for experimental interrogations, while evaluations at intermediate times expose quantal effects, such as those due to virtual excitations, that are not directly measurable in experiments. In this work, we adopt Eq. \eqref{eq:observable} to evaluate observables of the reaction system.

\section{Algorithm}
\label{sec:Algorithm}
In this section, we show our time-dependent quantum algorithm to simulate the subset of the nuclear inelastic scattering problems. We divide the discussion into seven parts: 1) the preparation of the initial state; 2) the approximation of the time-evolution operator; 3) the decomposition of the time-evolution operator in the basis representation; 4) the construction of the complete time evolution; 5) the construction of the quantum circuit; 6) the quantum simulation and measurement; and 7) the complexity of the tBFQ algorithm. In the following, we present the details of these topics.

\subsection{Preparation of the initial state}
\label{sec:initialStatePreperation}
For our protocol of quantum simulating the subset of the nuclear inelastic scattering problems, we first prepare the initial state of the reaction system. In this work, we map the basis states $\{ |\beta _j \rangle \}$ to a number of qubit configurations, which are formed by various sequences of binaries with each binary corresponding to one qubit in the quantum register. Following this compact encoding scheme, as an example, we have the mapping from the basis states $\{ |0 \rangle , \ |1 \rangle , \ |2 \rangle , \ |3 \rangle  \}$ to the two-qubit configurations as:
\begin{align}
|0 \rangle \mapsto | 00 \rangle , \ |1 \rangle  \mapsto | 10 \rangle , \ |2 \rangle  \mapsto | 01 \rangle , \ |3 \rangle  \mapsto | 11 \rangle .
\end{align}    
In general, it takes $ \lceil \log _2 N_{d} \rceil $ qubits, as the minimum, to represent the basis space of dimension $N_d$.

In the basis representation, the initial state vector of the reaction system, $|\psi ; t_0 \rangle $, is a superposition of the bases, as illustrated by Eq. \eqref{eq:fullyEntangledWF}. In the quantum simulations, the qubit configuration that corresponds to $|\psi ; t_0 \rangle $ can be created by applying a sequence of elementary gates, e.g., the Pauli-X and/or the Hadamard gate(s) \cite{Chuang:2000}, to the initial qubit configuration (usually defaulted as $|0 0 0 \cdots \rangle $) in the quantum register.

\subsection{Approximation of the time-evolution operator}
We start with the discretization of the scattering time. In particular, the scattering duration $[t_0,\ t_f]$ is discretized into $n$ partitions, with the equal time-step length to be $\Delta t = (t_f-t_0)/n$. The time-evolution operator [Eq. \eqref{eq:timeEvolutionOperator}] can then be approximated as 
\begin{align}
\hat{U} (t_f;t_0) \approx \hat{T} \Bigg\{ \exp \Big[ -i \big[ \hat{H}(t_f) \Delta t  + \cdots + \hat{H}(t_k ) \Delta t + \cdots + \hat{H}(t_{2}) \Delta t  + \hat{H}(t_1 ) \Delta t \Big]   \Bigg\}  .
\end{align}
Note that the accuracy of this approximation depends on the magnitude of $\Delta t$. In principle, we require  $||\hat{H}(t_i)|| \cdot \Delta t \ll 1$ for any moment during the scattering, i.e., $t_i \in [t_0,t_f] $.

We further approximate $ \hat{U} (t_f; t_0) $ by a series of unitary evolution operators according to the first-order Trotterization \cite{Chuang:2000}:
\begin{align}
\hat{U} (t_f;t_0) = { e^{-i \hat{H}(t_f) \Delta t} }  \cdots { e^{-i \hat{H}(t_k ) \Delta t} } \cdots { e^{-i \hat{H}(t_2) \Delta t} } { e^{-i \hat{H}(t_1) \Delta t} } + \mathcal{O}(\Delta t ^2) \label{eq:productFormula},
\end{align}
with $t_k$ being the discretized scattering time ($k=1,\ 2, \ \cdots , \ n$ and $t_n = t_f$). The accuracy of the first-order Trotterization is up to $(\Delta t )^2$ \cite{Chuang:2000}. In principle, a finer time-step improves the accuracy in the Trotterization at the cost of computational resources. Note that our approximation of $\hat{U} (t_f;t_0)$ can be improved by applying higher-order Trotterization formalism \cite{Suzuki:2005,Smith:2019}. As in Refs. \cite{Wu:2002,Sun:2020}, we take $  e^{-i \hat{H}(t ) \Delta t} $ to vary with the discretized scattering time, while every $  e^{-i \hat{H}(t_k ) \Delta t} $ holds the same for the time step $[t_{k-1}, t_k]$. 

For the subset of the nuclear inelastic scattering problems specified by the additivity and factorization assumptions in Eqs. \eqref{eq:summationOfInteraction} and \eqref{eq:factorizationOfInteractions}, we can further approximate the time-evolution operator for a single time step as \cite{Chuang:2000}
\begin{align}
e^{-i \hat{H}(t_k) \Delta t} = e^{- i ( \hat{H}_0 + \sum ^m _{i=1}  \hat{V}_i (t_k)  )\Delta t } = e^{-i \hat{H}_0 \Delta t} e^{-i \hat{V}_1(t_k ) \Delta t} e^{-i \hat{V}_2(t_k ) \Delta t} \cdots e^{-i \hat{V}_m(t_k ) \Delta t} + \mathcal{O}(\Delta t^2) . \label{eq:multipleProductMore}
\end{align}

\subsection{Decomposition of the time-evolution operator in the basis representation}
\label{sec:decompostionOfTrotterStep}
We start with the expression of the unitary $U(t_k)$, which is the matrix of the time-evolution operator for a single time step at $t_k$, $ e^{-i \hat{H}(t_k) \Delta t} $, in the basis representation. Indeed, the element of $U(t_k)$ can be obtained from Eq. \eqref{eq:multipleProductMore} by inserting the identity operators $\sum _j |\beta _j \rangle \langle \beta _j |$ as 
\begin{align}
\langle \beta _j | e^{-i \hat{H}(t_k) \Delta t} | \beta _l \rangle \approx & \sum _{pqr \cdots s} \langle \beta _j | e^{-i \hat{H}_0 \Delta t} | \beta _p \rangle \langle \beta _p | e^{-i \hat{V}_1(t_k) \Delta t} | \beta _q \rangle \langle \beta _q | e^{-i \hat{V}_2(t_k) \Delta t} | \beta _r \rangle \cdots \langle \beta _s | e^{-i \hat{V}_m(t_k) \Delta t} | \beta _l \rangle . \label{eq:multipleProductMatElement}
\end{align} 
In the matrix form, Eqs. \eqref{eq:multipleProductMore} and \eqref{eq:multipleProductMatElement} read 
\begin{align}
U(t_k) \approx U_{H_0,d} U_{V_1}(t_k) U_{V_2}(t_k) \cdots U_{V_m}(t_k) ,
\end{align}
where the element of $U_{H_0,d}$ is  $\langle \beta _j | e^{-i \hat{H}_0 \Delta t} | \beta _l \rangle $, and that of $U_{V_i}(t_k)$ is $ \langle \beta _j | e^{-i \hat{V}_i(t_k) \Delta t} | \beta _l \rangle $ (with $i=1,\ 2,\ \cdots , \ m$).

Next, we seek to decompose the unitary $U(t_k)$ such that 1) the time-dependent parts and the time-independent parts factorize; and 2) the time-dependent parts are directly parameterized according to the scattering time (or Trotter time). In doing so, the resulting unitaries are convenient for the quantum circuit construction. In the following, we discuss the formalism of the $U(t_k)$ decomposition leading to our result in Eq. \eqref{eq:unitaryForSingleTrotterStep}.

In particular, we proceed with the following prescriptions:
\begin{enumerate}

\item The matrix element of $U_{H_0,d}$ can be evaluated based on Eq. \eqref{eq:H0MatElement} as 
\begin{align}
\langle \beta _j | e^{-i \hat{H}_0 \Delta t} | \beta _l \rangle = e^{-i E_j \Delta t} \delta _{jl} . \label{eq:h0elementExplicit}
\end{align}
The unitary $U_{H_0,d}$ is hence diagonal\footnote{In this work, we use the subscript ``$d$" to denote diagonal matrix.} in the basis representation. We note that $U_{H_0,d}$ is independent of the scattering time. That is, $U_{H_0,d}$ is the same for all the time-evolution steps during the scattering.

\item The unitary $U_{V_i}(t_k)$ (with $i=1,\ 2,\ \cdots ,\ m $) can be evaluated based on the eigen-equation of the operator $\hat{W} _i$:
\begin{align}
\hat{W}_i | \zeta _{i \alpha} \rangle = {w} _{i \alpha}  | \zeta _{i \alpha} \rangle \label{eq:EigenEquationWi} ,
\end{align}
where ${w} _{i \alpha}$ is the eigenvalue that corresponds to the eigenbasis $| \zeta _{i \alpha} \rangle $, with $\alpha $ being the label. Since $\hat{W}_i$ is time independent [Eq. \eqref{eq:factorizationOfInteractions}], neither the eigenbasis set $\{ | \zeta _{i\alpha} \rangle \}$ nor the corresponding eigenvalue set $\{ w_{i \alpha} \}$ depends on the scattering time. The set $\{ | \zeta _{i\alpha} \rangle \}$ forms the time-independent eigen-representation of $\hat{W}_i$.

For each eigenvector $ | \zeta _{i\alpha} \rangle $, the following identity holds \cite{Arfken:2011}:
\begin{align}
e^ {-i \hat{V}_i (t_k) \Delta t } | \zeta _{i\alpha} \rangle = e^ {-i a_i (t_k) \hat{W}_i \Delta t } | \zeta _{i \alpha} \rangle = e ^{ -i a_i(t_k) w _{i \alpha} \Delta t } | \zeta _{i \alpha} \rangle  .
\end{align}	
The matrix element of the operator $ e^ {-i \hat{V}_i(t_k) \Delta t} $ in the basis representation can then be evaluated as
\begin{align}
\langle \beta _j | e^ {-i \hat{V}_i(t_k) \Delta t} | \beta _l \rangle = \sum _{\alpha , \gamma } \langle \beta _j | \zeta _{\alpha} \rangle \langle \zeta _{\alpha} | e^ {-i a_i(t_k) \hat{W}_i \Delta t} | \zeta _{\gamma } \rangle \langle \zeta _{\gamma } |  \beta _l \rangle = \sum _{\alpha}  \langle \beta _j | \zeta _{\alpha} \rangle e^{-ia_i(t_k){w} _{i \alpha} \Delta t } \langle \zeta _{\alpha} | \beta _l  \rangle  . \label{eq:matrixElementOfVi}
\end{align} 

Based on Eq. \eqref{eq:matrixElementOfVi}, the unitary $U_{V_i}(t_k)$ factorizes as:
\begin{align}
U_{V_i}(t_k) = U_i U_{i,d} (t_k) U_i^{\dagger} , \label{eq:DiagonalRepSpecialCase}
\end{align}
where the transformation matrix $ U_i $ is constructed from the amplitudes $ \{ \langle \beta _j  | \zeta _{i \alpha}  \rangle \} $, which characterizes the transformation between the basis representation and the eigen-representation of the operator $\hat{W}_i$. $ U_{i,d}(t_k) $ is a diagonal unitary with entries $\{ e^ { -i a_i(t_k) {w} _{i \alpha} \Delta t } \}$.\footnote{Note that, throughout this paper, we write explicitly the argument $t_k$ to denote those unitaries and gates that are parameterized by the scattering time. For example, in Eq. \eqref{eq:DiagonalRepSpecialCase}, $U_i$ (and the corresponding gate $\widetilde{U}_i$) does not depend on the scattering time (holds the same for the entire dynamics simulation), while $U_{i,d} (t_k) $ (and the corresponding gate $\widetilde{U}_{i,d}(t_k)$) is a function of the scattering time.}

\item 
According to Eqs. \eqref{eq:multipleProductMatElement}, \eqref{eq:h0elementExplicit} and \eqref{eq:DiagonalRepSpecialCase}, in the basis representation, the matrix of the time-evolution operator for a single time step at $t_k$ can be decomposed into a series of unitaries:
\begin{align}
U(t_k) \approx U_{H_0,d} \ \underbrace{ U_1 U_{1,d} (t_k) U_1^{\dagger} }_{U_{V_1}(t_k)} \ \underbrace{ U_2 U_{2,d} (t_k) U_2^{\dagger} }_{U_{V_2}(t_k)} \cdots \underbrace{ U_m U_{m,d} (t_k) U_m^{\dagger} }_{U_{V_m}(t_k)}
 \label{eq:unitaryForSingleTrotterStep} .
\end{align}
We note that $ U_i $ (with $i=1, \ 2, \ \cdots , \ m$) and $U_{H_0,d}$ are time independent. Once the set of eigenenergies $\{ E_j \}$, the basis sets $\{ |\beta _j \rangle \} $ and $ \{ | \zeta _{i \alpha} \rangle \}  $ are determined according to Eqs. \eqref{eq:EigenEquationH0} and \eqref{eq:EigenEquationWi}, $ U_i $ and $U_{H_0,d}$ are fixed for all the time-evolution steps during the scattering. 
On the other hand, we note that the diagonal unitary $ U_{i,d}(t_k) $ is time dependent. However, this time dependence is simple: it is determined only by $a_i(t_k)$ in the exponents of the entries of $ U_{i,d}(t_k) $, i.e., $\{ e^{-ia_i(t_k) w_{i\alpha} \Delta t} \}$. It is therefore easy to update $ U_{i,d}(t_k) $ according to the scattering time.  

\end{enumerate}

\subsection{The construction of the complete time evolution}
\label{sec:Sec3p4}
In essence, the decomposition shown in Eq. \eqref{eq:unitaryForSingleTrotterStep} provides a scattering-time-parameterized module to construct the unitary for every time-evolution step in the simulation of scattering dynamics. The unitary of the complete time-evolution operator $\hat{U} (t_f;t_0)$ can be calculated as a sequence of such modules according to Eq. \eqref{eq:productFormula}:
\begin{align}
U(t_f;t_0) = &  U(t_f) \cdots U(t_k) \cdots U(t_1) + \mathcal{O}(\Delta t^2) \nonumber \\ 
		  = & \underbrace{U_{H_0,d} \ U_1 U_{1,d} (t_f) U_1^{\dagger} \  U_2 U_{2,d} (t) U_2^{\dagger}  \cdots U_m U_{m,d} (t_f) U_m^{\dagger} } _{U(t_f)} \nonumber \\
		    & \times \cdots  \underbrace{U_{H_0,d} \ U_1 U_{1,d} (t_k) U_1^{\dagger} \  U_2 U_{2,d} (t_k) U_2^{\dagger}  \cdots U_m U_{m,d} (t_k) U_m^{\dagger} } _{U(t_k)} \nonumber \\	
		    & \times \cdots \underbrace{U_{H_0,d} \ U_1 U_{1,d} (t_1) U_1^{\dagger} \  U_2 U_{2,d} (t_1) U_2^{\dagger}  \cdots U_m U_{m,d} (t_1) U_m^{\dagger} } _{U(t_1)} + \mathcal{O}(\Delta t^2) . \label{eq:totalTimeEvolution}
\end{align}
We note that the complete time evolution for the scattering problem in the time-dependent external fields [represented by Eqs. \eqref{eq:summationOfInteraction}, \eqref{eq:factorizationOfInteractions} and \eqref{eq:timeEvolutionOperator}] does not involve classical computation at intermediate steps. That is, the tBFQ algorithm involves only quantum computation, without any hybrid computation at intermediate time steps.

\subsection{Quantum circuit construction}
\label{sec:CircuitConstruction}
\begin{figure}[H]
\centering
\includegraphics[width=13cm]{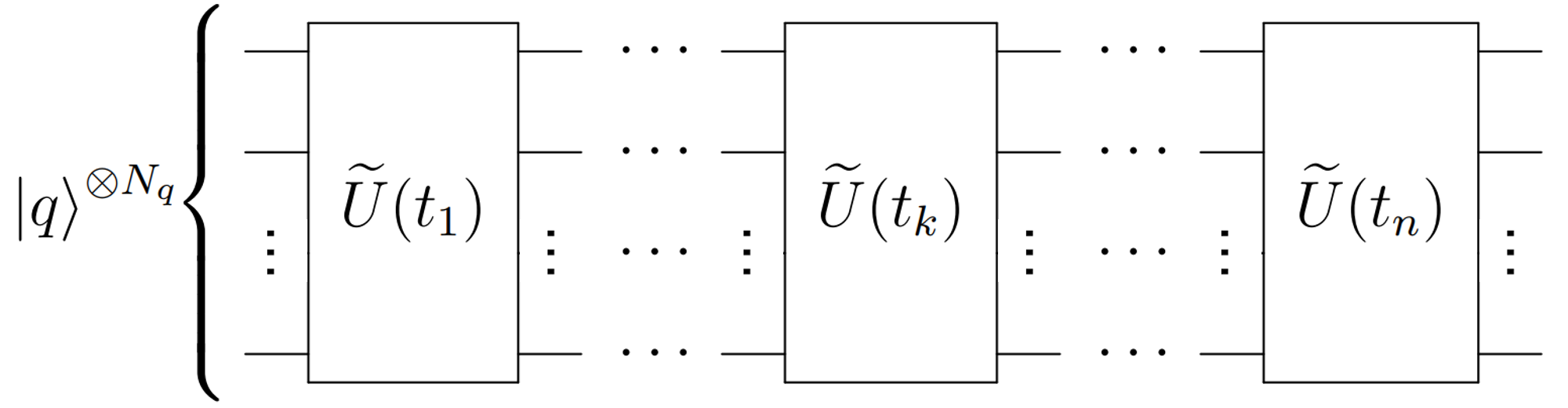}
\caption{Schematic quantum circuit design of $\widetilde{U}(t_f;t_0)$ for the complete time evolution according to Eq. \eqref{eq:totalTimeEvolution} on a $N_q$-qubit quantum register. The circuit operates from the left to right. The gate $\widetilde{U}(t_k)$, with $t_k=t_1,\ t_2,\ \cdots , \ t_n$ and $t_n = t_f$, corresponds to the time-evolution operator of a single time step at scattering time $t_k$. Though varying with $t_k$, the $\widetilde{U}(t_k)$ is considered to be time-invariant during each short time step $[t_{k-1},t_k]$. $\ket{q} ^{\otimes N_q}$ denotes the $N_q$-qubit register.}
\label{fig:totalEvolution}
\end{figure}

We follow a compact encoding scheme described in Sec. \ref{sec:initialStatePreperation} and qubitize the basis representation (see also in Sec. \ref{sec:simulationConditions} below for a detailed illustration). Following this compact encoding scheme, the number of qubits required for representing $N_d$ bases is $ \lceil \log _2 N_{d} \rceil $. 
We construct the quantum circuit based on this encoding scheme. For the purpose of illustration, we show in Fig. \ref{fig:totalEvolution} the schematic quantum circuit that corresponds to the time-evolution operator of the complete scattering $\hat{U}(t_f;t_0)$ [Eq. \eqref{eq:productFormula}]. We recall that the underlying idea of simulating $\hat{U}(t_f;t_0)$ is: 1) discretizing the scattering time into short time steps; and 2) approximating $\hat{U}(t_f;t_0)$ by the Trotter decomposition, where the resulting sequence of time-evolution operators $\hat{U}(t_k)$ ($t_k=t_1,\ t_2,\ \cdots , \ t_n$ and $t_n=t_f$) varies with scattering time $t_k$, while every $\hat{U}(t_k)$ is considered to be time-invariant on the very time step it acts. Correspondingly, in Fig. \ref{fig:totalEvolution}, the circuit of $\hat{U}(t_f;t_0)$ is constructed from a sequence of gates $\widetilde{U}(t_k)$, where each $\widetilde{U}(t_k)$ is taken to be the same throughout the short time step $[t_{k-1},t_k]$, while $\widetilde{U}(t_k) $ varies between different times steps in the simulation.

\begin{figure}[H]
\centering
\includegraphics[width=16cm]{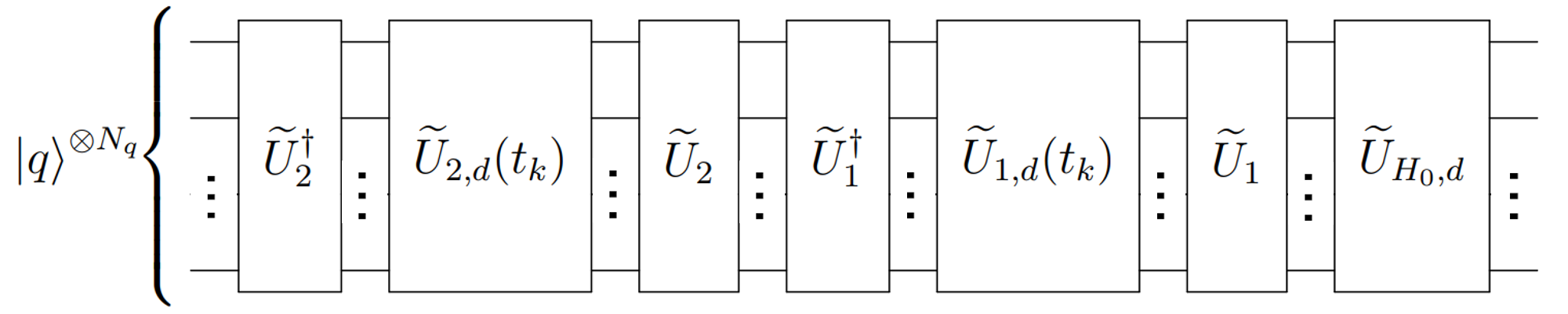}
\caption{Schematic circuit design for the time-evolution operator of a single time step, $\hat{U}(t_k)$, for the case with $m = 2$ on a $N_q$-qubit quantum register. $\ket{q} ^{\otimes N_q}$ denotes the $N_q$-qubit register. The circuit operates from the left to right. The circuit illustrates the idea to design the gate $\widetilde{U}(t_k)$ [see in Fig. \ref{fig:totalEvolution}] according to the decomposition principle Eq. \eqref{eq:unitaryForSingleTrotterStep}. More complicated cases (with larger $m$) can be generalized based on this design. See details in the text.
}
\label{fig:SingleTrotterStepOfModelProblem}
\end{figure}

Next, we construct the circuit to realize $\widetilde{U}(t_k)$, which evolves the reaction system for the time step $[t_{k-1},t_k]$. We achieve this by applying the decomposition principle in Eq. \eqref{eq:unitaryForSingleTrotterStep}, whereas each of the unitaries on the right-hand side of Eq. \eqref{eq:unitaryForSingleTrotterStep} corresponds to a specific quantum gate (e.g., the unitary $U_{H_0,d}$ corresponds to the gate $\widetilde{U}_{H_0,d}$). As a result, $\widetilde{U}(t_k)$ becomes the combination of gates $\widetilde{U}_i$, $\widetilde{U}_i^{\dagger}$, $\widetilde{U}_{H_0,d}$, and $\widetilde{U}_{i,d}(t_k)$ ($i=1,\ 2, \ \cdots , \ m$). For the purpose of illustration, we show the schematic circuit design of $ \widetilde{U}(t_k) $ for the case $m=2$ on a $N_q$-qubit quantum register in Fig. \ref{fig:SingleTrotterStepOfModelProblem}. Acting from the left and compared with Eq. \eqref{eq:unitaryForSingleTrotterStep}, the first 3 gates correspond to the unitary $ U_{V_2}(t_k) = U_2 U_{2,d}(t_k) U_2^{\dagger} $, while the second 3 gates correspond to the unitary $U_{ V_1}(t_k) = U_1 U_{1,d}(t_k) U_1^{\dagger}$. The rightmost gate corresponds to the unitary $U_{H_0,d}$. 

We can construct the circuit of $\widetilde{U}(t_k)$ such that it is of fixed structure with gate parameters being explicit functions of the scattering time. This can be seen from the components of $\widetilde{U}(t_k)$: 1) the unitaries ${U}_i$, ${U}_i^{\dagger}$ and ${U}_{H_0,d}$ are all time-independent; the corresponding gates $\widetilde{U}_i$, $\widetilde{U}_i^{\dagger}$ and $\widetilde{U}_{H_0,d}$ are hence fixed for all the time-evolution steps; and 2) the other type of unitary, ${U}_{i,d}(t_k)$ (diagonal), is explicitly scaled by the scattering time (see in the explanation below Eq. \eqref{eq:unitaryForSingleTrotterStep}); the corresponding $\widetilde{U}_{i,d}(t_k)$ can be realized with circuit of fixed structure with parameters determined by the scattering time (see, e.g., Appendix \ref{sec:Appendix} for a na\"ive approach).

$\widetilde{U}(t_k)$ serves as a module to construct the complete circuit $\widetilde{U}(t_f;t_0)$ for the scattering process. Since all such modules can be realized by simply tuning the scattering time $t_k$ with the circuit structure being the same, the quantum computer can automatically construct $\widetilde{U}(t_f;t_0)$ with the module $\widetilde{U}(t_k)$ without the aid from classical computers at intermediate steps during quantum simulation.

\subsection{Quantum simulation and measurement}
With the prescriptions described above, we construct the complete quantum circuit for the scattering. As illustrated in Fig. \ref{fig:totalEvolution}, the complete circuit consists of a sequence of modules sorted according to the scattering time, where 1)
these modules share the same scattering-time-invariant part (of which both the circuit structure and the gate parameters are the same); and 2) the remaining time-dependent part in each module is of fixed circuit structure with gate parameters being functions of the scattering time. Overall, the structure of a module repeats with only scattering time altered, such that the entire circuit for the dynamics simulation can be automatically generated on the quantum computer.

We then evolve the initial state according to the complete quantum circuit for simulating the scattering. At the end of each evolution, we measure all the qubits simultaneously and therefore collapse the wave function of the reaction system. After a set of such simulations, we collect the probabilities of basis states $\{ |C_{j}(t_f)|^2 \}$, which represent our knowledge of the reaction system at the end of the scattering. According to the probabilities $\{ |C_{j}(t_f)|^2 \}$, the observables of the reaction system can then be computed. 

We can also evolve the initial state to a fixed intermediate time $t_k$ during the scattering. The corresponding quantum circuit can be constructed by combining a sequence of scattering-time-parameterized modules up to the time $t_k$, according to Fig. \ref{fig:totalEvolution}. As such, the probabilities of basis states $\{ |C_{j}(t_k)|^2 \}$ and the quantity $\langle \hat{O}(t_k) \rangle $ at intermediate times of the scattering can also be obtained from the quantum simulations and measurements. We remark these values at the intermediate time during the scattering are quantal effects for tracking the dynamics and they are not experimentally measurable.

\subsection{The complexity of the tBFQ algorithm}
We focus on the applications of the tBFQ algorithm to a subset of the nuclear inelastic scattering problems that are of physical interest (see definitions in Sec. \ref{sec:Hdynamics}). For such problems, the time-dependent, full Hamiltonian ${H}(t)={H}_0+ \sum _{i=1} ^m {V}_i(t)$ is sparse and its norm $|| {H}(t) ||$ bounded for the entire scattering $t\in [t_0,t_f]$. In particular, within the basis representation, we find that 1) $H_0$ is diagonal with $|| H_0 ||$ bounded due to the importance truncation scheme applied to regulate the basis set (see discussions in Sec. \ref{sec:BasisRepresentation}); 2) each $ {V}_i(t) $ (with $i=1,\ 2, \cdots , \ m$) is a sparse matrix due to the selection rules of the transition operator \cite{Bohr1998}; and 3) for realistic scatterings, $ || {V}_i(t) || $ is bounded, while the total number of interaction channels/types $m$ is limited (e.g., taking $m=2$ to account for the low-energy Coulomb excitation processes in Sec. \ref{sec:ModelProblem} is sufficient).  

For the time-evolution step at arbitrary time $t_k$ during the scattering, the unitary of time evolution $U(t_k) $ in the basis representation can be decomposed into $3m+1$ unitaries according to Eq. \eqref{eq:unitaryForSingleTrotterStep}. We recall that 1) $U_{H_0,d}$ is diagonal (sparsity 1); 2) the transformation unitary $U_i$, together with the diagonal unitary $U_{i,d}(t_k)$ (sparsity 1), is solved from the spectral decomposition of the corresponding sparse matrix $V_i(t_k)= a_i(t_k) \cdot W_i(t_k)$ [Eq. \eqref{eq:EigenEquationWi}]. In this work, we assume that $U_i$ is both row and column sparse and take the sparsity of $U_i$ as $D_i$.\footnote{We define $D_i$ as the larger one of the row sparsity and the column sparsity of $U_i$.} Then, according to Ref. \cite{Jordan:2009,Aharonov:2003}, each one of these $3m+1$ unitaries can be simulated to precision $\epsilon $ using circuit of size bounded by $poly(\log _2 N_d, T, D_{\max}, 1/\Delta t , 1/\epsilon )$, where $N_d$ denotes the dimension of the basis space, and $T=t_f-t_0$ represents the total scattering duration. $D_{\max }$ is the maximal sparsity of all the transformation unitaries in Eq. \eqref{eq:unitaryForSingleTrotterStep}, i.e., $D_{\max }=\max _{i} \{ D_i \} $ with $ i=1, \ 2, \ \cdots , m$. Note that $D_{\max}$ is a constant that applies to all the time-evolution steps according to the analysis of Eq. \eqref{eq:unitaryForSingleTrotterStep}.

Altogether, we sum over the cost of simulating all the time-evolution steps and our approximation of the complete circuit size is bounded by $poly(\log _2 N_d, T, D_{\max}, 1/\Delta t , 1/\epsilon )$. On the other hand, the classical algorithms for solving such initial value problems, where the final state vector is obtained by sequential matrix-vector multiplications, the complexity can be approximated by $(N_d)^2$ (operations for a single time step of evolution) times the number of time-evolution steps $T/\Delta t$. By comparing the tBFQ and the classical algorithms, we would expect an exponential speedup for simulating complicated inelastic scattering problems.

\section{Model problem}
\label{sec:ModelProblem}

\begin{figure}[H]
\centering
\includegraphics[width=10cm]{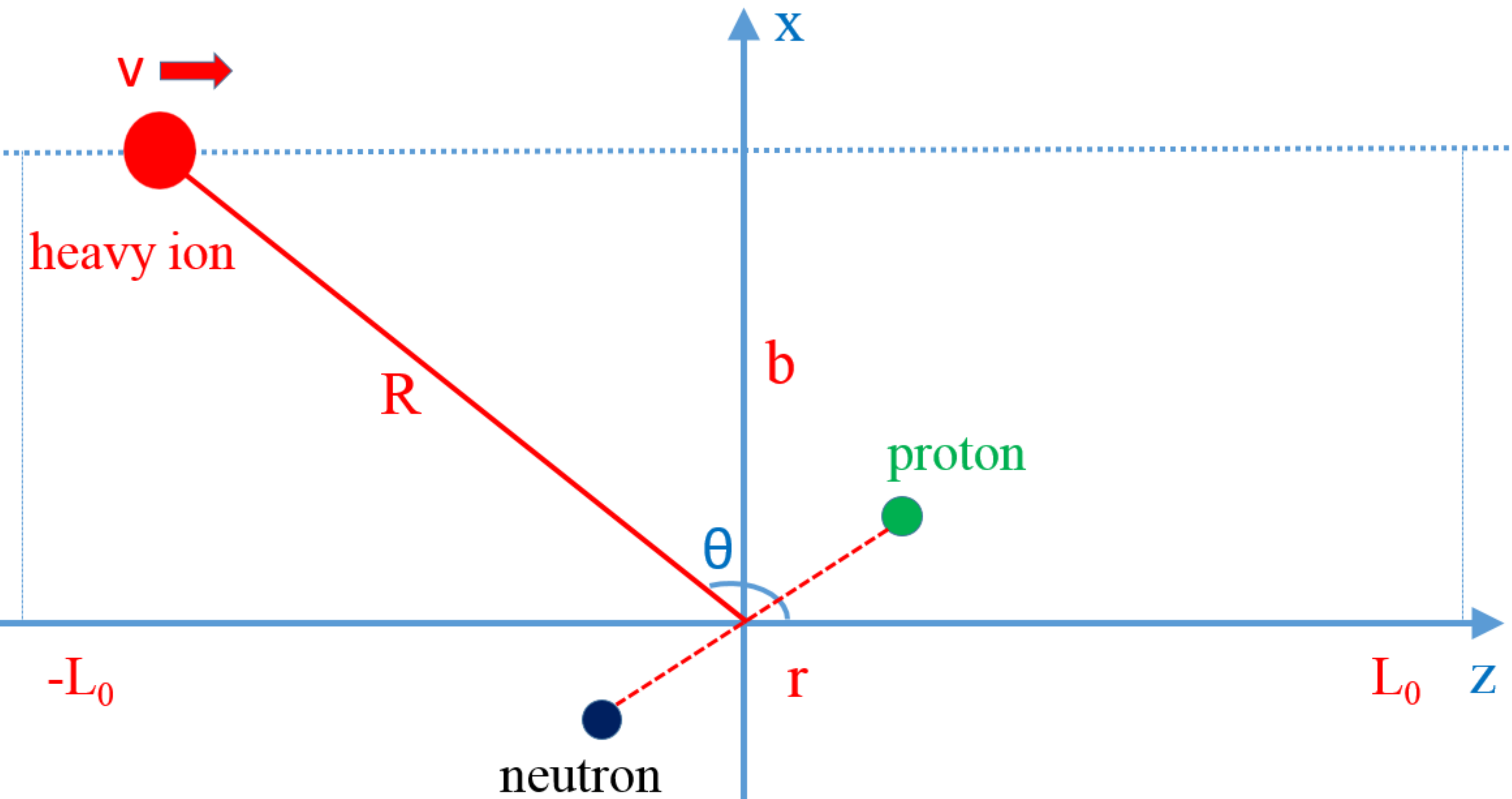}
\caption{(Color online) Setup for the Coulomb excitation of the deuteron in the peripheral collision with a heavy ion (adopted from Refs. \cite{Du:2018tce,Du:2017ckx}). See the text for more details.}
\label{fig:ScatteringSetup}
\end{figure}

We demonstrate our algorithm by applying it to a model problem: the Coulomb excitation of the deuteron in a peripheral collision with a heavy ion. In our previous papers, we introduced the time-dependent basis function (tBF) method and solved this test problem on classical computers \cite{Du:2018tce,Du:2017ckx}. Here, we only describe some of the necessary details to make our discussion self-contained.

Our setup of the model problem is shown in Fig. \ref{fig:ScatteringSetup}. We work in the center-of-mass frame of the deuteron target and take the scattering plane to be the $xz$ plane. For the purpose of illustration, we place the deuteron target in a weak harmonic oscillator trap. This trap regulates the continuum states and localizes the center-of-mass motion of the deuteron to simplify the problem. We assume that the neutron and proton are both point-like and the proton carries the unit charge $+e$. $\vec{r}$ denotes the position vector of the proton with respect to the neutron. The separation between the two nucleons is then $r\equiv |\vec{r}|$. The projectile is a heavy ion, which carries charge $Ze$ and travels with a constant velocity $\vec{v}$ parallel to the $\hat{z}$ axis. $\vec{R}$ denotes the position vector of the heavy ion from the origin. The impact parameter is $b$.

In this demonstration problem, we neglect nuclear interactions between the target and the projectile, and focus only on the deuteron target scattered by the external electromagnetic field produced by the heavy ion. Following our analyses in Refs. \cite{Du:2018tce,Du:2017ckx}, the total time-dependent Hamiltonian of the target is given by Eq. \eqref{eq:RefAndIntHamiltonians}, where $H_0$ is taken as the reference Hamiltonian that describes the intrinsic motion of the target:
\begin{align}
\hat{H}_0 = \hat{T}_{\rm rel} + \hat{V}_{\rm NN} + \hat{U}_{\rm trap} ,
\end{align}
with $\hat{T}_{\rm rel}$ being the relative kinetic energy of the neutron and the proton and $\hat{V}_{\rm NN}$ the nucleon-nucleon interaction. $\hat{U}_{\rm trap}$ denotes the external harmonic oscillator trap acting on the intrinsic degree of freedom of the target. 

As explained in Sec. \ref{sec:BasisRepresentation}, we employ $\hat{H}_0$ to solve for the eigenbasis set of the target $\{ | \beta _j \rangle \}$. While the deuteron can be solved easily by numerous techniques, we adopt the techniques of the No-Core Shell Model \cite{Barrett:2013nh, Navratil:2007we,Navratil:2000ww,Navratil:2000gs} anticipating the applications of this formalism for larger systems of nucleons. We neglect the excitations of the center of mass of the two-body system in this model application and focus on its internal excitations. In more realistic applications, the center of mass motion is not constrained by a trap and there is no effect of a trap on the intrinsic motion \cite{Yin:2019kqv}. 

At the limit of low incident speed $|\vec{v}|$, the external interaction $\hat{V}_{\rm int}(t)$ can be approximated by the dominant contribution from the electric dipole ($E1$) component of the time-varying Coulomb field \cite{Bohr1998,Alder:1956im}. In the basis representation, the external interaction in the Schr\"odinger picture reads:
\begin{align}
\langle \beta _j | \hat{V}_{\rm int} (t) | \beta _k \rangle = \frac{4\pi}{3}Ze^2 \sum ^{+1}_{\mu =-1} \frac{Y^{\ast}_{1\mu} (\Omega _R)}{|R(t)|^2} \underbrace{ \int d\vec{r} \langle \beta _j | \vec{r} \rangle \frac{r}{2} Y_{1\mu}(\Omega _r) \langle \vec{r} |\beta _k \rangle }_{\equiv \mathcal{I}(j,k,\mu)} , \label{eq:E1component}
\end{align}
where $Y_{\lambda \mu}$ denotes the spherical harmonics \cite{JSuhonen2007}: $\lambda =1 $ denotes the $E1$ component of the Coulomb field. The solid angles $\Omega _R$ and $\Omega _r$ are specified by the polar and azimuth angles of $\vec{R}$ and $\vec{r}$, respectively. Note that we factorize the interaction matrix element into the time-dependent part, which results from the time-varying external field, and the time-independent part, which comes from the intrinsic motion of the deuteron (i.e., $\mathcal{I}(j,k,\mu)$).

As for the current model problem, we sort the terms in Eq. \eqref{eq:E1component} according to $\mu $:
\begin{align}
\langle \beta _j | \hat{V}_{\rm int} (t) | \beta _k \rangle = \underbrace{ \sqrt{\frac{2\pi}{3}} Ze^2  \frac{\sin \theta (t)}{R^2(t)} }_{\equiv a_1(t)} \underbrace{ \Big[ \mathcal{I}(j,k, \mu =-1) - \mathcal{I}(j,k, \mu =+1) \Big] }_{\equiv \langle \beta _j |  \hat{W}_1 | \beta _k \rangle }  + \underbrace{ \sqrt{\frac{4\pi}{3}} Ze^2  \frac{\cos \theta (t)}{R^2(t)} }_{\equiv a_2(t)} \ \underbrace{ \mathcal{I} (j,k,\mu =0) }_{ \equiv \langle \beta _j |  \hat{W}_2 | \beta _k \rangle} , \label{eq:VintForTheModelProblem} 
\end{align}
where the trigonometric functions can be evaluated based on the scattering setup in Fig. \ref{fig:ScatteringSetup} as: 
\begin{align}
\sin \theta (t) = \frac{b}{R(t)} , \ \cos \theta (t) = \frac{-L_0 + v\cdot t}{R(t)} ,
\end{align}
with $R(t) = \sqrt{b^2 + (-L_0 + v\cdot t)^2}$. Note that we also factorize the time-dependent term $a_i(t)$ resulting from the time-varying external field and the time-independent term $ \langle \beta _j |  \hat{W}_i | \beta _k  \rangle $ that characterizes the intrinsic properties of the deuteron system. Eq. \eqref{eq:VintForTheModelProblem} can hence be rewritten as:  
\begin{align}
\langle \beta _j | \hat{V}_{\rm int} (t) | \beta _k \rangle = \langle \beta _j | \underbrace{ a_1(t) \hat{W}_1 }_{\equiv \hat{V}_1 (t)} | \beta _k \rangle + \langle \beta _j | \underbrace{ a_2(t) \hat{W}_2 }_{\equiv \hat{V}_2 (t)} | \beta _k \rangle .
\end{align}
We remark that, in the basis representation, the matrices $W_1$, $W_2$, and hence $V_{\rm int}(t) $ are all sparse due to the $E1$ selection rules. Since $H_0$ is diagonal, the full time-dependent Hamiltonian of the model problem can then be sorted into 3 sparse matrices as:
\begin{align}
H(t) = H_0 +  a_1(t) W_1 + a_2(t) W_2 .
\end{align}

As discussed in Sec. \ref{sec:Algorithm}, in the basis representation, the unitary of the time-evolution operator for the time step $[t_{k-1},t_k]$ can be computed/decomposed as:
\begin{align}
U(t_k) \approx U_{H_0,d} \ \underbrace{ U_1 U_{1,d} (t_k) U_1^{\dagger} }_{U_{V_1}(t_k)} \ \underbrace{ U_2 U_{2,d} (t_k) U_2^{\dagger} }_{U_{V_2}(t_k)}
 \label{eq:ModelProblemunitaryForSingleTrotterStep} .
\end{align}
As explained above, the unitaries $U_1$, $U^{\dag}_1$, $U_2$, $U^{\dag}_2$ and $U_{H_0,d}$ are fixed for all the time-evolution steps according to the choice of the basis set $\{ |\beta _j \rangle  \}$ and the eigenenergies $\{ E_j \}$ of the reference Hamiltonian $H_0$ together with that of the eigenbasis $\{ |\zeta _{i\alpha} \rangle  \}$ of each operator $W_i$. On the other hand, $U_{1,d} (t_k)$ and $U_{2,d} (t_k)$ are directly parameterized according to the scattering time. The unitary $U(t_k)$ and its decomposition serve as an elementary, scattering-time-parameterized module to construct the unitary of the complete time-evolution operator. 

The quantum circuit for an arbitrary time-evolution step can be designed according to Eq. \eqref{eq:ModelProblemunitaryForSingleTrotterStep}. The so-constructed quantum circuit is of fixed structure with gate parameters being explicit functions of scattering time, as shown in Fig. \ref{fig:SingleTrotterStepOfModelProblem}. We note that this circuit can be implemented as a scattering-time-parameterized module to construct the complete/partial circuit for simulating the complete/partial scattering process, as shown in Fig. \ref{fig:totalEvolution}.

\section{Simulation conditions}
\label{sec:simulationConditions}

\begin{table}[H]
\centering
\caption{Selected basis states of the deuteron target in the model problem. The target is confined in an external harmonic oscillator trap of strength 5 MeV and the center-of-mass motion of the target is restricted to the lowest state of the trap. The first and second columns present the quantum numbers of the selected states. The third and fourth columns present, respectively, the eigenenergy and the {\it r.m.s.} point-charge radius of each state: they contribute to the intrinsic energy and {\it r.m.s.} point-charge radius of the target according to Eq. \eqref{eq:observable}. The last column shows the 3-qubit configurations corresponding to these selected states.}   
\begin{tabular}{c c c c c} \hline \hline 
Channel                  & Magnetic substate & $ \langle E \rangle $ [MeV] & $ \langle r^2 \rangle ^{\frac{1}{2}}  $ [fm]  & Qubit configuration \\ \hline 
\multirow{3}{*}{$(^3S_1,^3D_1)$} & 	$M=-1$	   & \multirow{3}{*}{$-0.65289$}     & \multirow{3}{*}{$1.47222$}   & $|000 \rangle $ \\
                         &          $M=0$      &      				             &     & $|100 \rangle $ \\
                         &          $M=+1$     &                                  &    & $|010 \rangle $ \\ \hline
\multirow{1}{*}{$^3P_0$} & 			$M=0$ 	   & \multirow{1}{*}{$12.0733$}  & \multirow{1}{*}{$3.13427$}    & $|110 \rangle $ \\ \hline
\multirow{3}{*}{$^3P_1$} & 			$M=-1$	   & \multirow{3}{*}{$12.7585$}   & \multirow{3}{*}{$3.27644$}   & $|001 \rangle $ \\
                         &          $M=0$      &                              &     & $|101 \rangle $ \\
                         &          $M=+1$     &                               &    & $|011 \rangle $ \\  \hline \hline
\end{tabular}
\label{tab:SevenBases}
\end{table}

We take the projectile to be a bare (all electrons removed) uranium nucleus, $U^{92+}$, which is treated as a point-like source of the time-varying external Coulomb potential. The incident speed is set to be 0.1 in units of the speed of light, while the impact parameter is chosen to be $5$ fm. We also take the exposure time to be from -3 to 3 MeV$^{-1}$, which corresponds to a time duration of approximately $4.0 \times 10^{-21}$ sec. That is, the time evolution starts when the uranium is about 60 fm before its closest approach to the origin, and concludes when the uranium is in the position 60 fm after the closest approach. Beyond this region, the induced intrinsic excitations of the target are negligible since the Coulomb field is sufficiently weak compared with the available excitations.

We choose the same basis representation as in our previous works Refs. \cite{Du:2018tce,Du:2017ckx}. In particular, we obtain the basis states (or reaction channels) of the target $\{ | \beta _j \rangle \}$ according to Eq. \eqref{eq:EigenEquationH0}. For the purpose of demonstration, we trim the basis size and retain only the lowest seven states in three interaction channels: $(^3S_1, ^3D_1)$, $^3P_0$, and $^3P_1$. These states are listed in Table \ref{tab:SevenBases}, based on which we construct the basis representation for our model problem. Our basis representation can be directly mapped to a qubit representation, as provided in the last column in Table \ref{tab:SevenBases}. 


We construct the quantum circuit for our model problem based on the scattering-time-parameterized modules, as discussed in Sec. \ref{sec:ModelProblem}. For this exploratory work, we decompose the relevant gates in the module as follows.\footnote{We remark that improving the circuit design, e.g., by reducing the circuit depth and the CNOT-gate counts, will be an effort of future work. Our focus here is to illustrate the methodology of simulating nuclear inelastic scattering on quantum computers.} We apply the column-by-column decomposition scheme (see Ref. \cite{Iten:2016} and references therein) to obtain the circuits of the time-invarying gates $\widetilde{U}_1$, $\widetilde{U}_2$, $\widetilde{U}^{\dag}_1$, $\widetilde{U}^{\dag}_2$, and $\widetilde{U}_{H_0,d}$ from the respective unitaries ${U}_1$, ${U}_2$, ${U}^{\dag}_1$, ${U}^{\dag}_2$, and ${U}_{H_0,d}$ in Eq. \eqref{eq:ModelProblemunitaryForSingleTrotterStep}. In practice, we apply the UniversalQCompiler package \cite{Iten:2019} to obtain these circuits. As for the time-dependent gates $\widetilde{U}_{1,d}(t_k)$ and $\widetilde{U}_{2,d}(t_k)$, we obtain the corresponding circuits by decomposing the respective diagonal unitaries $U_{1,d}(t_k)$ and $U_{2,d}(t_k)$ following the prescriptions in Refs. \cite{Barenco:1995,Iten:2016}. In particular, for the current 3-qubit model problem, we present the details of our preliminary decomposition scheme in Appendix \ref{sec:Appendix}.

We choose the initial state of the deuteron system in our simulations to be $(^3S_1, ^3D_1),M=-1$, which corresponds to the $|000 \rangle $ state of the quantum register. We take the evolution-time step to be $\Delta t = 0.01$ MeV$^{-1}$. We simulate our model problem on an ideal quantum simulator provided by IBM Qiskit \cite{Santos:2017,Cross:2017}. That is, we do not consider the limitations of the real-world quantum computers, such as the gate noise and the decoherence effect, in this work.

\section{Results and discussions}
\label{sec:resultsAndDiscussions}
\begin{figure}[h]
\centering
\includegraphics[width=16.5cm]{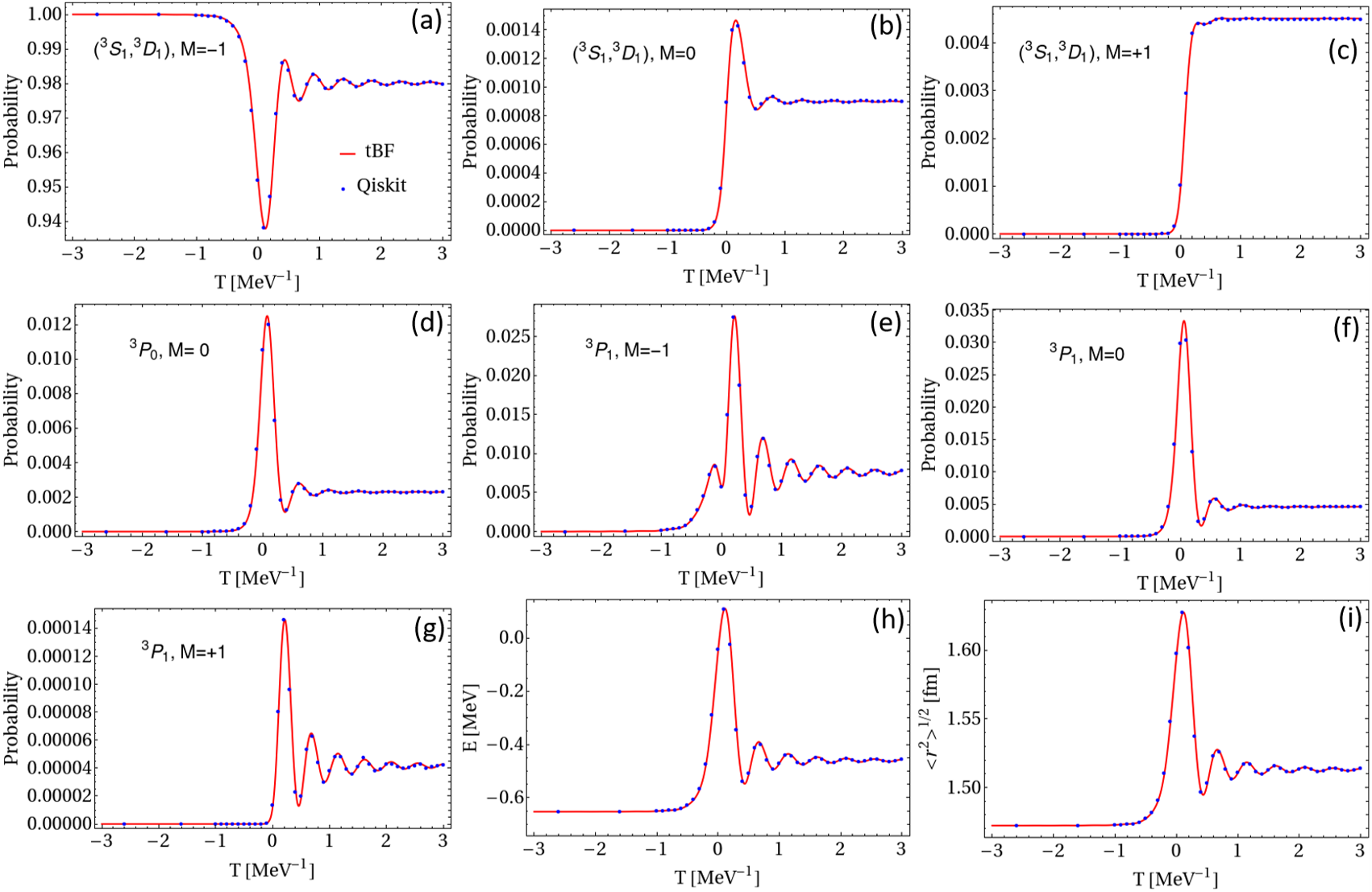}
\caption{(Color online) Transition probabilities and scattering observables calculated by the tBF method and the tBFQ method. Panels (a)-(g) [panel (a) presenting the initial state] are the transition probabilities of the selected basis states in Table \ref{tab:SevenBases}. Panels (h) and (i) show the excitation of intrinsic energy and the expansion of the state-average {\it r.m.s.} point-charge radius of the deuteron system arising from the Coulomb excitation, where the results are obtained by both the classical tBF method (smooth red lines in each panel) and the tBFQ method via quantum simulations on IBM Qiskit (discrete blue points in each panel). 
}
\label{fig:ProbabilityComparison}
\end{figure}

The transition probabilities of all the retained channels (Table \ref{tab:SevenBases}) are presented in panels (a)-(g) of Fig. \ref{fig:ProbabilityComparison}. For comparison, we also show the results computed based on a classical algorithm, the time-dependent basis function (tBF) \cite{Du:2018tce}. We find good agreement between the results calculated by the quantum and classical algorithms.\footnote{
As a crosscheck, we also calculate the ``Trotterized” evolution via the classical approach. This is achieved by sequentially multiplying matrices of dimension $2^3 \times 2^3$ according to Eq. \eqref{eq:totalTimeEvolution}. We find that the classical Trotterization results agree well with the tBF results (within the error of less than 1$\%$ for each state). Since in panels (a)-(g) of Fig. \ref{fig:ProbabilityComparison}, the curves of the tBF and the classical Trotterization results overlap with each other, we do not present explicitly the classical Trotterization results.
}

We work with the ideal quantum simulator. Besides the error from the approximation of the first-order Totterization, which is at the order of $(\Delta t)^2$, there is the expected variance from quantum mechanical measurement, which is statistical. For the current work, we take 10 quantum simulations, each with $10^7$ measurements, for every chosen evolution time shown in Fig. \ref{fig:ProbabilityComparison} (note that the number of measurements per simulation can be adjusted according to the requirement of the simulation precision; we can reduce the number of measurements per simulation for a lower precision requirement). The resulting statistical errors are less than $1 \%$ for most of the retained states, except for $^3P_0, M=0$ and $^3P_1, M=0$, where such errors are less than $4 \%$. We remark that these statistical errors are expected to decrease with increasing number of simulations. In Fig. \ref{fig:ProbabilityComparison}, we do not show the error bars explicitly since they are smaller than the size of the symbols in the plots. 

In Fig. \ref{fig:ProbabilityComparison}, we also present the results of the two selected observables, i.e., the intrinsic energy [panel (h)] and the state-average {\it r.m.s.} point-charge radius of the deuteron target [panel (i)] throughout the Coulomb excitation process.\footnote{We remark that the state-average {\it r.m.s.} point-charge radius of the deuteron target is measured with respect to the center-of-mass so this radius is 1/2 of the separation $ r $ defined in Fig. \ref{fig:ScatteringSetup}.} We obtain these results according to Eq. \eqref{eq:observable}. As explained in Sec. \ref{sec:BasisRepresentation}, we note that only the asymptotic values of these observables (i.e., when the Coulomb field is sufficiently weak at the end of the scattering) are subject to experimental interrogations, while the intermediate results obtained in the middle of the scatterings (when the external field is effective and the intrinsic transitions take place) are quantal effects which are not experimentally measurable. As may be expected from the good agreement in the state population obtained from the quantum and the classical methods, the corresponding results of the observables also show a good agreement. Working with the ideal quantum computer simulator, we note that the error bars of the results from the quantum simulations come mainly from the statistics in quantum measurement. These error bars are smaller than the size of the symbols and are hence omitted.

Finally, we remark that the quantum evolution according to the tBFQ algorithm is fully coherent. That is, the wave function of the deuteron evolves in a fully coherent way during the simulation until it collapses in measurement. This is one of the appealing aspects of Feynman’s original idea \cite{Feynman:1981tf} of simulating one quantum system via another one, in which the full coherence (and also entanglement) is preserved naturally. In addition, tBFQ reveals the non-perturbative features of the scattering process. By keeping only the dominant $E1$ multipole component of the Coulomb field, one would expect populations in the states $^3P_0, M=0$; $^3P_1, M=-1$; and $^3P_1, M=0$, according to the $E1$ selection rule [recall that we prepare the initial state to be $(^3S_1,^3D_1), M=-1$]. However, tBFQ keeps all the higher-order effects (e.g., sequential $E1$ excitations) as well, which contribute to a complicated transition network among all the retained states. As a result, this network feeds the ``$E1$-forbidden" states that can not be directly populated from the initial state, such as $(^3S_1,^3D_1), M=+1$, during the Coulomb excitation.

\section{Conclusion and Outlook}
\label{sec:FinalSec}
We present the time-dependent quantum algorithm for the basis function treatment of nuclear inelastic scattering on qubits (tBFQ). This algorithm employs the output from classical nuclear structure calculations (e.g., eigenbasis set and eigenenergies). It provides an approach for quantum simulating a subset of the nuclear inelastic scattering problems, where the internal degrees of freedom of the reaction system are excited by the external interaction.

For tBFQ, we work in the Schr\"odinger picture and divide the full Hamiltonian of the reaction system undergoing scattering into the reference Hamiltonian (presumably time independent), which determines the available excitations of that system, and the time-dependent external interaction, which drives the dynamical excitation processes. For the subset of the nuclear inelastic scattering problems, we assume that the external interaction can be divided according to a sum of (dominant) terms, where each of the terms couples the time-dependent part from the external actions (e.g., fields) and the time-independent part that acts on the intrinsic degrees of freedom of the reaction system.  

We solve for the eigenbases of the reference Hamiltonian and apply an importance truncation to reduce the size of the basis set. This trimmed basis set is used to construct the basis representation, within which we solve the full Hamiltonian and the time-evolution operator. We qubitize the basis representation via a compact encoding scheme and design the quantum circuit that is directly parameterized according to the scattering time based on the time-evolution operator. According to the measurements of the simulations, we obtain the transition probabilities (which in turn determine the inelastic scattering cross section) and the other observables of the reaction system undergoing scattering. For the subset of the inelastic scattering problems that are of physical interest, we expect that the tBFQ algorithm would achieve an exponential speedup in simulating complicated scattering problems which classical algorithms would find intractable. 

For illustrative purposes, we demonstrate this algorithm with a model problem, the Coulomb excitation of the deuteron in a peripheral collision with a heavy ion. The results of the transition probabilities and the selected observables computed by the tBFQ algorithm applying the IBM Qiskit quantum simulator agree well with the corresponding results computed by classical methods.

The promise of quantum computers is that the computational complexity of many-body Hamiltonian dynamics can be exponentially reduced. However, many challenges exist. Going forward, the optimization of the algorithm, e.g., constructing efficient quantum circuits \cite{Berry:2015,Berry:2014}, will be necessary for simulations of general reaction problems on real quantum computers. The general features of our algorithm also allow its use for studying many-body Hamiltonian dynamics in atomic and subatomic physics \cite{Bates:2012}.

\section*{Acknowledgments}
This work was supported in part by the US Department of Energy (DOE) under Grants No. DE-FG02-87ER40371, No. DE-SC0018223 (SciDAC-4/NUCLEI), and No. DESC0015376(DOE Topical Collaboration in Nuclear Theory for Double-Beta Decay and Fundamental Symmetries). A portion of the computational resources were provided by the National Energy Research Scientific Computing Center (NERSC), which is supported by the US DOE Office of Science. 
X.Z. and W.Z. are supported by the Strategic Priority Research Program of Chinese Academy of Sciences, Grant No. XDB34000000. 
X. Z. is supported by new faculty startup funding by the Institute of Modern Physics, Chinese Academy of Sciences, by Key Research Program of Frontier Sciences, Chinese Academy of Sciences, Grant No. ZDB-SLY-7020, and by the Natural Science Foundation of Gansu Province, China, Grant No. 20JR10RA067. W.Z. are supported by the National Natural Science Foundation of China (Grants No. 11975282 and No. 11435014) and the Key Research Program of the Chinese Academy of Sciences (Grant No. XDPB15).
We acknowledge P. Yin and Y. Li for their valuable input. W.D. and J.P.V. thank H. Lamm, G. R. Luecke, R. Basili and D. Zhao for valuable discussions.

\newpage

\begin{appendices}

\section{Decomposition of diagonal unitary}
\label{sec:Appendix}
In this section, we follow Refs. \cite{Barenco:1995,Iten:2016} and present our preliminary prescription to decompose a diagonal unitary matrix of size $2^3 \times 2^3 $ with time varying entries into the corresponding quantum circuit.  
To be specific, the unitary matrix is of the form 
\begin{align}
{P}(t) = 
\bordermatrix{
	      & 000  & 001  & 010 & 011 & 100 & 101 & 110 & 111 \cr
      000 & e^{-ia(t)} &  &  &   &   &  &   &     \cr
      001 &  & e^{-ib(t)} &  &   &   &  &   &     \cr
      010 &  &  & e^{-ic(t)} &   &   &  &   &    \cr
      011 &  &  &  & e^{-id(t)}  &   &  &   &    \cr
      100 &  &  &  &   & e^{-ie(t)}  &  &   &    \cr
      101 &  &  &  &   &   & e^{-if(t)} &   &     \cr
      110 &  &  &  &   &   &  & e^{-ig(t)}  &     \cr
      111 &  &  &  &   &   &  &   &  e^{-ih(t)}    \cr
      }      \label{eq:matrixDecomposition}  ,  
\end{align}
where $a(t)$, $b(t)$, $c(t)$, $d(t)$, $e(t)$, $f(t)$, $g(t)$, and $h(t)$ are real functions of time $t$. To proceed, we label the columns and rows by the binary configurations $000$, $001$, $\cdots $, $111$. We take the leftmost/middle qubit to be the most/second significant qubit and apply them to control the least significant (rightmost) qubit.

We decompose ${P}(t)$ as the circuit below
$$
\Qcircuit @C=1em @R=1em @!R {
& \ctrlo{1}   & \qw   & \ctrlo{1} & \qw   & \ctrl{1}   & \qw  & \ctrl{1}   & \qw  \\
& \ctrlo{1}   & \qw   & \ctrl{1} & \qw   & \ctrlo{1}   & \qw  & \ctrl{1}   & \qw  \\
& \gate{ \widetilde{U}_a} & \qw   & \gate{\widetilde{U}_b} & \qw & \gate{\widetilde{U}_c} & \qw  & \gate{\widetilde{U}_e} & \qw 
}
$$
where the uppermost/middle/lowermost line denotes the most/second/least significant qubit. In the matrix form, these gates are
\begin{align}
& U_a(t) =
\begin{pmatrix}
e^{-ia(t)} &  \\ 
 & e^{-ib(t)}
\end{pmatrix} , \ 
U_b(t) =
\begin{pmatrix}
e^{-ic(t)} &  \\ 
 & e^{-id(t)}
\end{pmatrix} , \\
& U_c(t) =
\begin{pmatrix}
e^{-ie(t)} &  \\ 
 & e^{-if(t)}
\end{pmatrix} , \ 
U_e(t) =
\begin{pmatrix}
e^{-ig(t)} &  \\ 
 & e^{-ih(t)}
\end{pmatrix} . 
\end{align}

The decomposition of the Eq. \eqref{eq:matrixDecomposition} can then be done with the following circuit identity \cite{Barenco:1995,Iten:2016}
$$
\Qcircuit @C=1em @R=1em @!R {
& \ctrl{1} & \qw & & & \qw & \qw      & \qw      & \qw      & \ctrl{2} & \ctrl{1} & \qw             & \ctrl{1} & \qw      & \gate{ \widetilde{E} } & \ctrl{2}   & \qw      & \qw  & \\
& \ctrl{1} & \qw & \push{\rule{.3em}{0em}=\rule{.3em}{0em}} & & \qw & \gate{ \widetilde{E} } & \ctrl{1} & \qw      & \qw      & \targ    & \gate{\widetilde{E}^{\dag}} & \targ    & \ctrl{1} & \qw      &   \qw        & \qw      & \qw  &\\
& \gate{\widetilde{U}_x} & \qw & & & \qw & \gate{\widetilde{C} } & \targ    & \gate{\widetilde{B}} & \targ    & \qw      & \gate{\widetilde{B}^{\dag}} & \qw      & \targ    & \gate{\widetilde{B}} & \targ      & \gate{\widetilde{A}} & \qw  & 
}
$$
where $\widetilde{U}_x$ is a general unitary gate. $\widetilde{A}$, $\widetilde{B}$, $\widetilde{C}$ are special unitary gates and $\widetilde{E}$ is the phase gate (see Ref. \cite{Barenco:1995} for details). As a special case in this work, we consider 
\begin{align}
U_x(t) =
\begin{pmatrix}
e^{-ix(t)} &  \\ 
 & e^{-iy(t)}
\end{pmatrix} ,
\end{align}
where $x(t)$ and $y(t)$ are real functions on time $t$. Then the gates in the above circuit identity are
\begin{align}
& \widetilde{A}= \widetilde{I}, \\
& \widetilde{B}(t) = \widetilde{R}_z[-\delta(t)], \ \widetilde{B}^{\dag}(t) = \widetilde{R}_z[\delta(t)], \\ 
& \widetilde{C}(t) = \widetilde{R}_z[\delta(t)], \ \widetilde{C}^{\dag}(t) = \widetilde{R}_z[-\delta(t)] , \\
& \widetilde{E}(t) = |0\rangle \langle 0| + e^{i\alpha(t)}| 1 \rangle \langle 1 |, \ \widetilde{E}^{\dag}(t) = |0\rangle \langle 0| + e^{-i\alpha(t)}| 1 \rangle \langle 1 |, 
\end{align}
with the time-dependent parameters $\delta(t) = \frac{x(t)-y(t)}{4}$ and $\alpha(t) = - \frac{x(t)+y(t)}{4} $. 
$\widetilde{I}$ denotes the identity gate, while $\widetilde{R}_z[\theta]$ is the rotational gate about the $\hat{z}$ axis with the angle $\theta $ being the parameter. In the matrix form, these gates are  \cite{Chuang:2000}
\begin{align}
I =
\begin{pmatrix}
1 &  \\ 
 & 1
\end{pmatrix} , \ 
R_z[\theta] =
\begin{pmatrix}
e^{-i\theta /2} &  \\ 
 & e^{i\theta /2}
\end{pmatrix} , \ 
E(\alpha) =
\begin{pmatrix}
1 &  \\ 
 & e^{i \alpha }
\end{pmatrix} .
\end{align}

Since a gate controlled on $|1 \rangle $ can be transformed into a gate controlled on $|0 \rangle $ using two Pauli $\widetilde{ \sigma } _x$ gates \cite{Iten:2016}, e.g., 
$$
\Qcircuit @C=1em @R=1em @!R {
& \ctrl{1} & \qw & & & \qw & \ctrl{1} & \qw  & \qw \\
& \ctrlo{1} & \qw & \push{\rule{.3em}{0em}=\rule{.3em}{0em}} & & \gate{\widetilde{ \sigma } _x} & \ctrl{1} & \gate{\widetilde{ \sigma } _x} & \qw  \\
& \gate{\widetilde{U}_x} & \qw & & &  \qw & \gate{\widetilde{U}_x} & \qw & \qw
}
$$
the other parts of the matrix ${P}(t)$ can be decomposed following the procedures described above.

\end{appendices}

\end{document}